\documentclass[ reprint, superscriptaddress, floatfix, amsmath,amssymb, aps, prx ]{revtex4-2}

\usepackage[utf8]{inputenc}
\usepackage{amsmath,amssymb}
\usepackage{wrapfig}
\usepackage{graphicx}
\usepackage{bm}
\usepackage[colorlinks=true,linkcolor=blue,citecolor=blue,urlcolor=blue]{hyperref}
\usepackage{makecell}
\usepackage{siunitx}
\usepackage{cases}

\newcommand{\citeasnoun}[1]{Ref.~\cite{#1}}

\newcommand{\Figref}[1]{Figure~\ref{fig:#1}}
\newcommand{\figref}[1]{Fig.~\ref{fig:#1}}
\renewcommand{\eqref}[1]{Eq.~(\ref{eq:#1})}
\newcommand{\Eqref}[1]{Equation~(\ref{eq:#1})}

\newcommand{\vect}[1]{\boldsymbol{\mathbf{#1}}}

\newcommand{\secref}[1]{Sec.~\ref{sec:#1}}

\newcommand*{\Ev}{\vect{E}}
\newcommand*{\psiv}{\bm{\psi}}
\newcommand*{\Gv}{\vect{G}}
\newcommand*{\Bv}{\bm{\mathrm{B}}}
\newcommand*{\Iv}{\bm{\mathrm{I}}}
\newcommand*{\fv}{\bm{\mathrm{f}}}
\newcommand*{\Einc}{\Ev_{\rm inc}}

\newcommand*{\rv}{\vect{r}}

\newcommand*{\Pv}{\vect{P}}

\begin{document}
\preprint{APS/123-QED}

\title{Fullwave design of cm-scale cylindrical metasurfaces via fast direct solvers}

\author{Wenjin Xue}
\thanks{equal contribution}
\affiliation{Department of Electrical Engineering, Yale University, New Haven, Connecticut 06511, USA}
\affiliation{Energy Sciences Institute, Yale University, New Haven, Connecticut 06511, USA}
\author{Hanwen Zhang}
\thanks{equal contribution}
\affiliation{Energy Sciences Institute, Yale University, New Haven, Connecticut 06511, USA}
\affiliation{Department of Applied Physics, Yale University, New Haven, Connecticut 06511, USA}
\affiliation{Department of Mathematics, Yale University, New Haven, Connecticut 06511, USA}
\author{Abinand Gopal}
\affiliation{Department of Mathematics, Yale University, New Haven, Connecticut 06511, USA}
\author{Vladimir Rokhlin}
\affiliation{Department of Mathematics, Yale University, New Haven, Connecticut 06511, USA}
\author{Owen D. Miller}
\affiliation{Energy Sciences Institute, Yale University, New Haven, Connecticut 06511, USA}
\affiliation{Department of Applied Physics, Yale University, New Haven, Connecticut 06511, USA}

\date{\today}

\begin{abstract}
Large-scale metasurfaces promise nanophotonic performance improvements to macroscopic optics functionality, for applications from imaging to analog computing. Yet the size scale mismatch of centimeter-scale chips versus micron-scale wavelengths prohibits use of conventional full-wave simulation techniques, and has necessitated dramatic approximations. Here, we show that tailoring ``fast direct'' integral-equation simulation techniques to the form factor of metasurfaces offers the possibility for accurate and efficient full-wave, large-scale metasurface simulations. For cylindrical (two-dimensional) metasurfaces, we demonstrate accurate simulations whose solution time scales \emph{linearly} with the metasurface diameter. Moreover, the solver stores compressed information about the simulation domain that is reusable over many design iterations. We demonstrate the capabilities of our solver through two designs: first, a high-efficiency, high-numerical-aperture metalens that is 20,000 wavelengths in diameter. Second, a high-efficiency, large-beam-width grating coupler. The latter corresponds to millimeter-scale beam design at standard telecommunications wavelengths, while the former, at a visible wavelength of \SI{500}{nm}, corresponds to a design diameter of \SI{1}{cm}, created through full simulations of Maxwell's equations.
\end{abstract}

\maketitle
\section{Introduction}
Simulating Maxwell's equations underpins the exploration of wave phenomena across nanophotonics, from complex interference effects~\cite{Sol2023,Sauvan2013,Ganapati2013} to exotic band topology~\cite{Zhou2018,Cerjan2019,Yuan2021}. Yet for the emerging field of metasurfaces~\cite{Yu2014,Kamali2018,Lalanne2023}, where diameters are readily centimeter-scale (cm-scale) or larger, standard tools such as finite difference~\cite{jin2015theory,oskooi2010meep} and finite element~\cite{jin2015theory,reddy2019introduction} methods cannot provide accurate solutions in a reasonable time. One typically must employ dramatic approximations, such as the unit-cell approximation~\cite{khorasaninejad2015achromatic,chen2018broadband,wang2018broadband}, for proof-of-principle demonstrations. In this paper, we develop a ``fast direct'' solver, based on well-conditioned integral formulations of electromagnetic scattering, to achieve fast and accurate fullwave simulation and design of large-scale metasurfaces. Fast-direct solver techniques have been proposed and developed in the applied mathematics community over the past decade~\cite{martinsson2020fast,Chen2002,Gillman2011,Gillman2013,corona2015n,Ambikasaran2016,gopal2022accelerated}, descendent from fast multipole methods~\cite{greengard1987fast,greengard1988rapid,carrier1988fast,greengard1997new,cheng2006wideband}; we show that a modified fast-direct implementation tailored to the unique geometry of metasurfaces can lead to accurate simulations with computation times scaling \emph{linearly} (technically, quasilinearly) with the diameter of the metasurface. We implement the fast-direct technique for cylindrical metasurfaces (translation-invariant in one direction) of both TE and TM polarizations, and demonstrate superior scaling compared with a finite-difference approach. The speed and accuracy of this approach enables a striking demonstration: the fullwave design (by adjoint optimization) of a high-NA metalens that is \emph{twenty thousand} wavelengths in diameter. For a \SI{500}{nm} visible wavelength, this corresponds to a metasurface diameter of \SI{1}{cm}. Our simulation technique readily accepts any source field, which we illustrate in a second demonstration of a high-efficiency, large-area, aperiodic grating coupler. Our results showcase the promise of fast-direct fullwave solvers for large-area metasurfaces, enabling the full complexity of nanophotonic wave physics to be harnessed at macroscopic scales.

The workhorse simulation tools in nanophotonics are finite-difference and finite-element methods~\cite{oskooi2010meep,jin2015theory,reddy2019introduction}. There are a few advantages of these methods that have made them so popular. First, matrix discretizations of differential equations are relatively easy to implement. Second, the matrices are sparse, thanks to the local nature of electromagnetic interactions, such that solutions of the linear equations can easily be accelerated~\cite{Davis2006}. Finally, for a subset of these approaches---including the finite-difference time-domain (FDTD) method---parallelization can be efficient and relatively straightforward to implement. However, these collective advantages also come with drawbacks. Differential operators are unbounded (typically exhibited as eigenvalues going to infinity), leading to ill-conditioned matrix representations~\cite{duan2009high,martinsson2020fast}. Finite-resolution sampling creates wave-speed inaccuracies that cause increasingly large phase-accumulation errors at increasing scattering-region sizes. Hence, even if the amount of time to run an FDTD simulation nominally increases linearly with the size of a domain (for a fixed resolution), to keep the error fixed one has to increase the resolution, quickly leading to impractically large computation times for large scatterers.

Integral equations are a well-known alternative to their differential counterparts, arising through consideration of the current/polarization response fields within a scattering body (or at its interfaces) and enforcing consistency between these currents and the fields they excite. The central operator in an integral equation is a convolutional Green's-function operator that represents the field excited from a point source in the \emph{absence} of the scattering body. Green's functions encode the correct speed of light, preventing the phase error plaguing differential approaches. Moreoever, integral equations can be numerically well-conditioned by the nature of integral operators~\cite{martinsson2020fast}. These advantages, however, come with their own drawbacks. First, the singular nature of Green's functions (e.g., divergent fields at a point source) requires careful discretization. Second, the resulting matrices are dense, which can quickly lead to impossible long simulation times without clever acceleration techniques. A common approach for acceleration is to exploit the convolutional nature of Green's functions in an iterative linear-equation solver, which relies on fast matrix-vector multiplication.

Two leading integral-equation techniques are the discrete dipole approximation (DDA)~\cite{purcell1973scattering,draine1988discrete,draine1993beyond,draine1994discrete,yurkin2007discrete,yurkin2011discrete} and the method of moments (MoM)~\cite{chew2001fast,jin2015theory}. Their approaches highlight some of the tensions in overcoming the drawbacks discussed above. Both DDA and MoM techniques are typically implemented with iterative solvers that rely on fast matrix-vector multiplication. In DDA this is particularly simple, as the integral equation is discretized on a uniform grid whose translational symmetry enables use of fast Fourier transforms. However, it has traditionally been difficult to accurately discrete the singular Green's functions on a uniform grid, and DDA solutions tend to offer a low accuracy that further degrades with modest refractive indices or scatterer sizes~\cite{yurkin2007discrete}. Conversely, MoM methods typically use a reasonable numerical integration (quadrature) scheme, but incur significant complexity in generating triangular/tetrahedral meshes and the corresponding matrix elements, and then in developing algorithms for fast matrix-vector multiplication without the translational symmetry of the uniform grid. Both DDA and MoM solvers can achieve quasilinear computation times per iteration~\cite{chew2001fast}, $O(N)$ for $N$ degrees of freedom (``order $N$,'' up to logarithmic corrections), but the number of iterations needed for convergence increases with the scatterer size, ultimately stalling these approaches before they reach the largest possible sizes.

``Fast direct solvers''~\cite{martinsson2020fast} represent a promising new approach to solving integral equations. A noteworthy recent example is given in \citeasnoun{gopal2022accelerated}, which starts with a simple uniform grid, but uses the high-order quadrature scheme developed in \citeasnoun{duan2009high} to accurately resolve the Green's-function singularity. Then, instead of using an iterative solver that suffers at large scatterer sizes, \citeasnoun{gopal2022accelerated} use a direct solver build from a multilevel compression of the Green's function matrix, exploiting the information-sparsity of long-range interactions~\cite{martinsson2020fast}. This ``fast-direct'' solver successfully reduces the computation time of two-dimensional simulations from cubic scaling with $N$ simulation degrees of freedom, i.e., $O(N^3)$ (for standard dense linear algebra solvers) to $O(N^{3/2})$. Ideally, however, one would like a simulation approach that scales linearly with the problem size, in order to enable simulations at the largest size scales.

Complementary to such ``full-wave'' solvers are approximation techniques, exploiting physical simplifications to reduce computational complexity. For metasurfaces, for example, unit-cell~\cite{khorasaninejad2015achromatic,wang2018broadband,chen2018broadband} and overlapping-domain~\cite{lin2019overlapping} approximation methods are popular schemes for building large-area, single-layer devices from small-region constituents. Similarly, for waveguide and grating-coupler devices, adiabaticity is used to simplify in-plane propagation and out-of-plane coupling~\cite{johnson2002adiabatic,perez2018sideways,pestourie2018inverse}. However, in each case the set of devices that satisfy the requisite approximations is strongly constrained, and much of the multiple-scattering physics of electromagnetic waves must be discarded.

In this work, we improve the fast-direct, integral-equation-based simulation technique of \citeasnoun{martinsson2020fast} to efficiently simulate metasurface geometries at extraordinarily large scales (tens of thousands of wavelengths in diameter). We start by reviewing fast-direct solvers and describing our proposed improvements (\secref{FDSolvers}). Our fast solver exploits the large aspect ratio of metasurface geometries and has a simulation time that scales linearly with metasurface diameter. We demonstrate its capabilities in \secref{Examples}, where we test its simulation prowess against a standard FDTD simulation tool, and then use it design metasurfaces that are thousands of wavelengths in diameter. Finally, we conclude with a brief summary and discussion of new avenues to pursue (\secref{Conclusion}).

\section{A fast-direct solver for metasurfaces}
\label{sec:FDSolvers}
For a scattering domain $V$ with material susceptibility distribution $\chi(\vect{r})$, within which an incident field $\Einc$ induces a polarization-field response $\Pv(\vect{r})$, the volume integral (Lippmann--Schwinger) equation is
\begin{align}
    \Pv(\vect{r}) = \chi(\vect{r}) \left(\Einc(\vect{r}) + \int_V{\Gv(\vect{r},\vect{r'}) \Pv(\vect{r'})d\vect{r'}}\right),
    \label{eq:VIE}
\end{align}
where $\Gv$ is the background Green's function, whose convolution with $\Pv$ determines the scattered field. Throughout, we will assume a background of free space. Integral-equation methods can be adapted to non-vacuum backgrounds, though care is needed to ensure that the simulations remain fast and efficient. \Eqref{VIE} applies to any nonmagnetic scatterer in three dimensions, if $\chi$ is interpreted as a tensor field and $\Gv$ as a dyadic Green's function. For the remainder of this article, for simplicity of notation we will assume an isotropic scalar susceptibility in two dimensions. Generalizations to tensor susceptibilities are simple and straightforward. Generalizations to three-dimensional Green's functions are possible and, as discussed in the Conclusions section, represent an important future step.

In this section, we describe a fast-direct solver for \eqref{VIE}, tailored to metasurface geometries, i.e., scatterers with large aspect ratios. Starting from a square grid discretization, which is easy to set up but low accuracy, local correction terms (by ``Duan--Rokhlin quadrature'') can increase accuracy with only a few near-diagonal corrections (\secref{DRquad}). This Green's function can then be compressed by interpolative decomposition, using a binary tree structure that further aids in computing a complete or approximate inverse of the integral-equation matrix (\secref{FastSolvers}). If the approximate inverse is built, then a few iterations will quickly converge to the final solution. 

\subsection{Duan--Rokhlin quadrature}
\label{sec:DRquad}
Square grids are powerfully simple for discretization. They are especially common in finite difference methods~\cite{jin2015theory,oskooi2010meep}, where they yield relatively simple second-order-accurate approximations of differential operators. For integral equations, however, a difficulty arises for a square-grid discretization (or \emph{any} ``Nystr\"{o}m'' discretization, in which the unknown variables are the field values at specific positions, instead of basis-function coefficients) of \eqref{VIE}: the diagonal elements of the matrix representation of the Green's function operator would be proportional to the self-term $\Gv(\rv,\rv)$, which diverges. The simplest approach, which yields the DDA method, is to ignore the self term in the summation. In a two-dimensional scattering domain with discrete points $\rv_i$ on the grid, and grid spacing $h$, the DDA approximation of \eqref{VIE} is:
\begin{align}
\Pv(\rv_i)  = \chi(\rv_i) \left(\Einc (\rv_i) + \sum_{j\neq i} \Gv(\rv_i,\rv_j) \Pv(\rv_j) h^2 \right),
  \label{eq:DDA_eq}
\end{align}
which can be understood mathematically as the application of the ``punctured trapezoidal rule'' to \eqref{VIE}. As resolution increases and $h$ decreases, the error of the solution of \eqref{VIE} decays at best as $h^2$. Various versions of DDA approximate the self-term through physical polarizability arguments~\cite{purcell1973scattering,draine1988discrete,draine1993beyond,draine1994discrete,yurkin2007discrete,yurkin2011discrete}, but none of these approaches change the underlying error scaling as a function of resolution. The sizeable errors in the DDA approximation, and their relatively slow scaling with $h$, ultimately limit the size of the scattering body (e.g., metasurface) that can be simulated quickly and efficiently.

There is a simple-to-implement correction to \eqref{DDA_eq} that can dramatically increase its accuracy, without sacrificing the simplicity of the square-grid discretization. The idea was developed in \citeasnoun{duan2009high}, based on the following observation. Since a polarization field $\Pv$ is compactly supported and smooth for a compactly supported smooth scatterer (as discussed above), any integration of $\Pv$ by standard trapezoidal rule will converge super-algebraically, i.e., it will incur an error smaller than $h^m$ for any $m>0$ (cf. \citeasnoun{trefethen2022numerical}). Hence, the errors in \eqref{DDA_eq} as an approximation of \eqref{VIE} come entirely from the singular elements of $\Gv$, along the diagonal. In \citeasnoun{duan2009high}, such errors are systematically corrected by modifying a few weights near the diagonal. The first correction that one can make is to the singularity (diagonal) coefficient itself; for 2D scalar waves, correcting this term leads to $\sim h^4$ error scaling. Correcting successive neighbors by moment matching leads to higher-order accuracies. For a set $N_i$ that includes the point $\rv_i$ and its nearest neighbors (and possibly next nearest neighbors, etc.), one arrives at a modified version of \eqref{DDA_eq}:
\begin{align}
\Pv(\rv_i)  = \chi(\rv_i) \biggl(  \Einc(\rv_i) &+ \sum_{j\neq i}{\Gv(\rv_i,\rv_j) \Pv(\rv_j)h^2}  \nonumber\\
  & +  \sum_{\rv_j\in N_i}{\tau(\rv_i,\rv_j) \Pv(\rv_j)h^2}  \biggr)
  \label{eq:Duan_Rokhlin_eq}
\end{align}
\Eqref{Duan_Rokhlin_eq} successfully achieves both simple, square-grid discretization, as well as high accuracy. \citeasnoun{duan2009high} demonstrated up to \emph{tenth}-order accuracy for TE (scalar) waves with a 25-point stencil for $\tau(\rv_i,\rv_j)$; we find below that even fourth-order accuracy leads to significant improvements over conventional DDA, and requires only a single correction, along the diagonal, for each source point. The straightforward corrections of \eqref{Duan_Rokhlin_eq} highlight the more general idea that it is easier to achieve stable, high-order discretizations for integral equations than for their differential counterparts because numerical integration is free of numerical cancellations that plague numerical differentiation. The order of accuracy of a discretization of \eqref{VIE} is solely determined by the order of the numerical quadrature for convolving the Green's function against the polarization field~\cite{Colton2013}.

The simplicity of the above high-order quadrature scheme comes from the assumption that the scatterer is \emph{smooth}, i.e., without discontinuous boundaries. This might seem useless for the typical scenario in nanophotonics, with discrete materials and sharp boundaries. However, any discontinuous material interface can be well-approximated by a narrow but smooth transition. One can smooth susceptibility distributions using Planck-taper, sigmoid, and other filter functions, tailored to the discretization resolution to ensure that the boundary is sufficiently sampled, to obtain results close enough to exact solutions after applying quadrature corrections. 

By combining the quadrature scheme and boundary smoothing described above, we arrive at a matrix form of \eqref{VIE}, the volume integral equation:
\begin{align}
    (\bm{{\rm I}}-\bm{{\rm B G}})\bm{{\rm \psi}} = \bm{{\rm f}}
    \label{eq:matrix_eq}
\end{align}
where $\Iv$ is an identity matrix, $\Bv$ is a diagonal matrix with elements $\Bv_{i,i} = \chi (\rv_i) h^2$, $\bm{{\rm G}}$ is the free-space Green's function matrix, including correction terms $\tau$, and $\bm{{\rm \psi}}$ and $\bm{{\rm f}}$ are the vector representations of $\Pv$ and $\chi \Einc$, respectively. The next question, then, is how to efficiently \emph{solve} the matrix equation of \eqref{matrix_eq}. It is well-conditioned (due to its integral-equation origins), but the Green's-function matrix $\Gv$ is dense. A fast solver is needed to compute $\psiv = (\Iv - \Bv\Gv)^{-1} \fv$.


\subsection{Fast direct solver techniques}
\label{sec:FastSolvers}

\begin{figure*}[htb]
  \includegraphics[width=\textwidth]{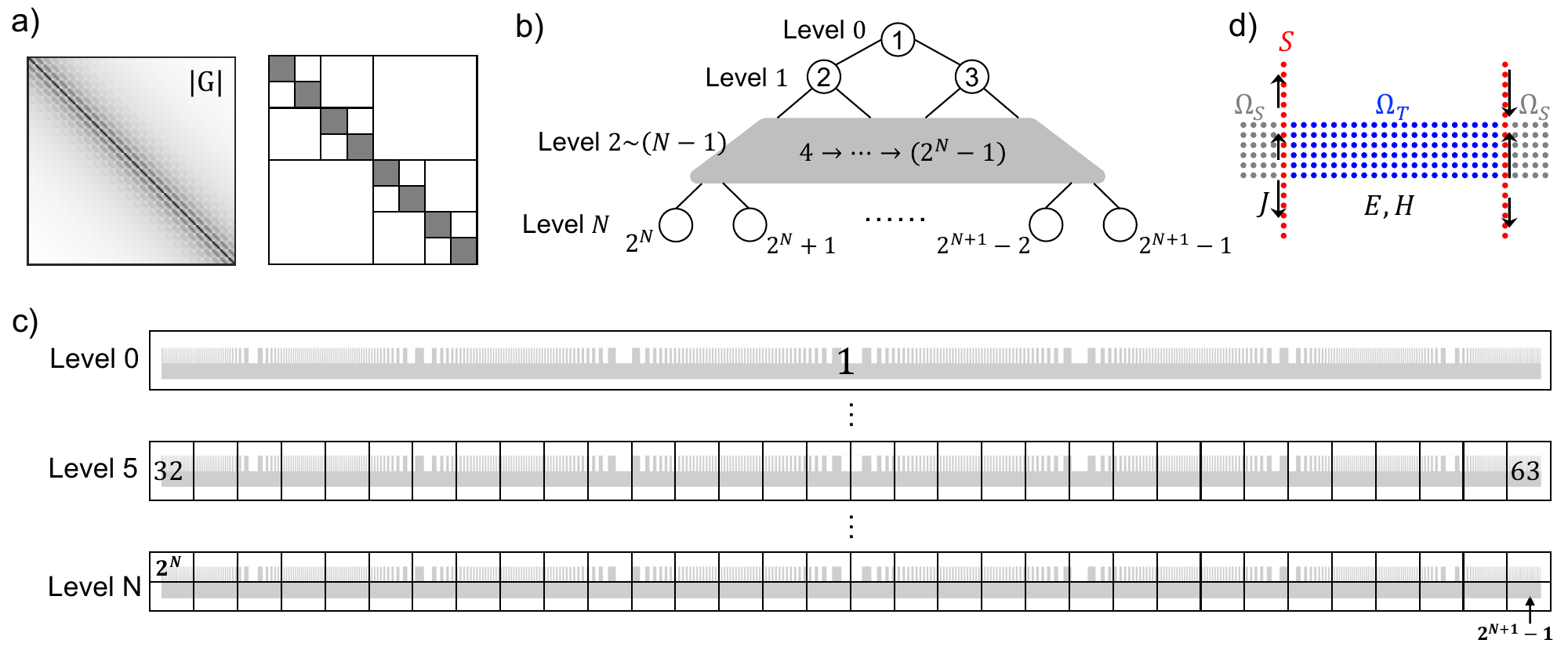}
  \caption{(a) Left panel: the magnitude of Green's function matrix $\Gv$, a dense matrix with slowly decaying off-diagonal components. Right panel: the HODLR rank structure of $\Gv$. The hierarchical off-diagonal components (white) are low-rank, leaving only a small number of small, full-rank diagonal blocks (gray). (b) A binary tree structure for non-recursive storage of $\Gv$. (c) Hierarchical partition of a metasurface simulation region. Each partitioned box corresponds to a node in the binary tree. (d) A tailored vertical-cut proxy surface to most efficiently compress the interactions of one volume of the scatterer with another. Usage of this proxy surface in tandem with compression and acceleration techniques leads, up to logarithmic corrections, to a computation time scaling linearly with diameter.}
  \label{fig:fig1}
\end{figure*}

Fast direct solvers compute the solution $\psiv=(\Iv-\Bv \Gv)^{-1} \fv$ directly and efficiently based on compressed representations of $\Gv$~\cite{gopal2022accelerated}. A low-rank matrix is easy to compress: one can simply store the singular vectors corresponding to nontrivial singular values. Unfortunately, $\Gv$ is neither low rank nor approximately low rank. However, it is \emph{hierarchically off-diagonal low rank} (HODLR), which means that the off-diagonal blocks of $\Gv$ are low rank, at every scale. For example, writing $\Gv$ as a block 2$\times$2 matrix, the two off-diagonal blocks are low rank. Moreover, each of the diagonal blocks can be decomposed into smaller 2$\times$2 blocks, each of which have off-diagonal blocks which themselves are low rank. And so on, recursively. This property is depicted in \figref{fig1}(a), which shows both the magnitude of the elements of a prototype Green's function, and the recursive decomposition of a Green's-function matrix into diagonal (shaded) and low-rank (white) blocks. Physically, the HODLR property arises because even though wave amplitudes decay slowly in space, the information contained in those amplitudes decays quickly~\cite{martinsson2020fast}. (For example, accurate modeling of the radiation into a fixed volume $V$ can require significantly fewer multipole moments as the separation increases.)

A HODLR representation can yield simple recursive expressions for the needed matrix inverse, but recursive computation can be inefficient in practice, and especially difficult to efficiently implement in a parallel computing environment. A non-recursive approach is to store each constituent component of the matrix in a binary tree structure, as in \figref{fig1}(b). Physically, this corresponds to physical subdivisions of a scatterer, as depicted in \figref{fig1}(b). A compressed representation of $\Gv$ can consist of the information of each block (of varying sizes) in \figref{fig1}(a). The diagonal (shaded) blocks require dense storage, but can all be relatively small. The off-diagonal blocks can be stored with any decomposition (singular value, interpolative, etc.) that exploits the low-rank nature of those blocks. The total storage scales as $O(kN\log{N})$, where $k$ is the upper limit of the off-diagonal rank of $\Gv$~\cite{martinsson2020fast}. In two dimensions, $k$ is proportional to $N^{1/2}$, while in three dimensions, $k\sim O(N^{2/3})$. Then matrix multiplication and matrix inversion can operate directly on these stored representations, in times that scale as $O(kN\log{N})$. For our simulations, we compress $\Gv$ into hierarchically block separable (HBS) form~\cite{gopal2022accelerated}, and we use interpolative decomposition (ID) for the low-rank approximation~\cite{gopal2022accelerated,martinsson2020fast}, as it saves compression time and memory. 

Further compression savings can typically be made by exploiting the surface equivalence principle: the fields emanating from a volume of currents can be exactly reproduced by appropriate surface currents. Hence, the low-rank approximation of the interaction $\Gv$ between any two blocks in \figref{fig1}(c) can replace the source volume with discrete surface points (on that volume), which is referred to as a ``proxy surface''~\cite{gopal2022accelerated,martinsson2020fast}. By utilizing a number of points proportional to the surface instead of the volume, the compression complexity can be reduced, with a compression time that scales os $O(N^{3/2})$ instead of $O(N^2)$~\cite{gopal2022accelerated}.

However, for metasurface geometries that are thin in the direction perpendicular to their diameter, with enormous aspect ratios, the surface area to volume ratio is much larger than for a box domain, rendering such proxy surfaces ineffective in reducing computation time. We introduce a new twist on the proxy-surface approach, specifically tailored to metasurfaces. In a metasurface, at the first few tree levels (the larger domains in \figref{fig1}(b)), the dominant field interaction is typically horizontal wave propagation, such that discretizing the surfaces with equal number of points along top/bottom sidewalls as their left/right counterparts is wasteful (and leads to artificial inflation of the apparent rank of the interaction). Instead, we propose using an \emph{open} proxy surface, vertical cuts extending above and below the metasurface, as depicted in \figref{fig1}(d), for these large-domain, first tree levels, which unlocks faster compression via proxy surfaces in metasurface geometries. 

\begin{figure*}[htb]
  \includegraphics[width=\textwidth]{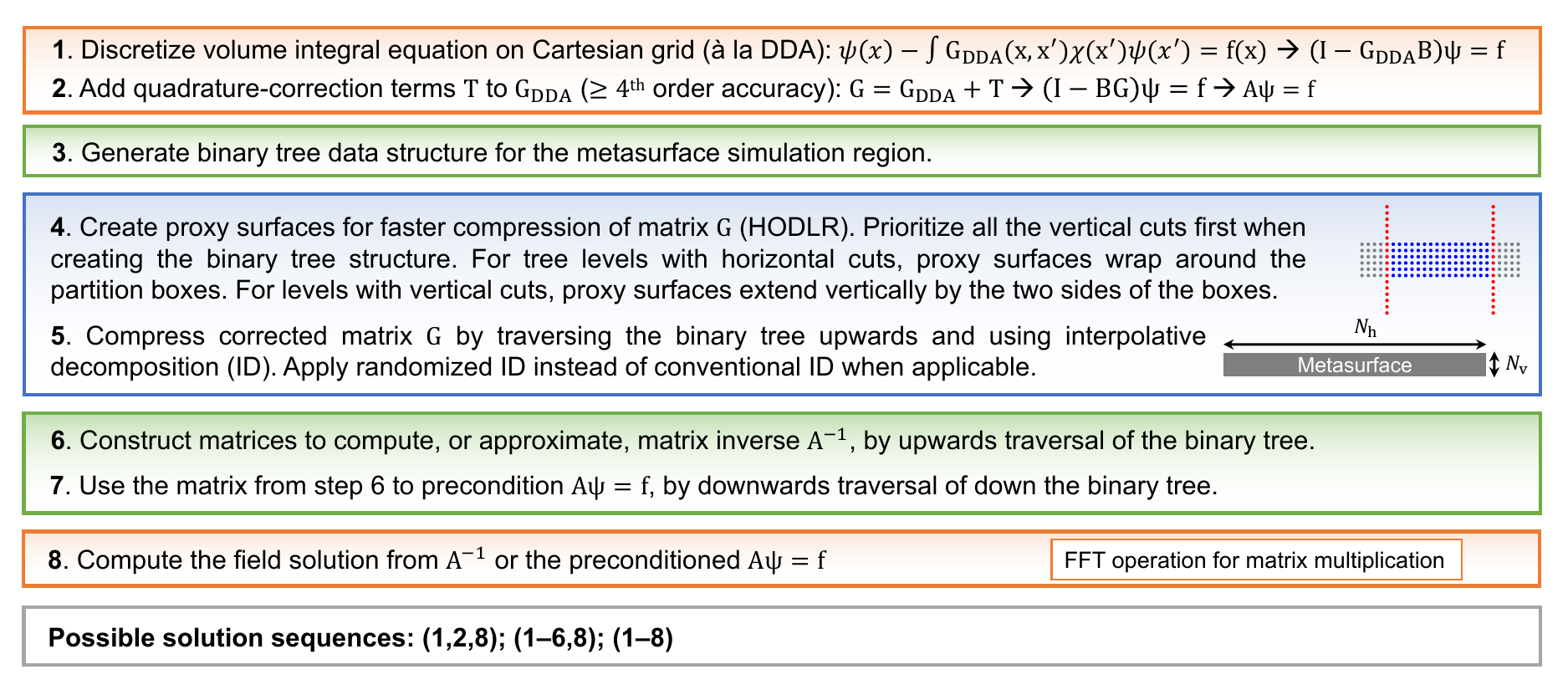}
  \caption{Procedure for fast and accurate integral-equation-based metasurface simulations. The orange boxes outline the high-order quadrature method developed in \citeasnoun{duan2009high}, while the green boxes highlight fast-direct solver techniques from \citeasnoun{martinsson2020fast}. The blue box highlights the metasurface-adapted proxy surface and randomized ID adaptations that we propose, enabling linear-in-diameter scaling of the simulation time. Smaller structures can be simulated with iterative methods (steps 1,2,8), but the largest structures require compression and a direct solver (steps 1--6,8) or an approximate direct solve with a few ``polish'' iterative solves at the end (steps 1--8).}
  \label{fig:fig2}
\end{figure*}

Given the HBS compression of $\Gv$, exploiting our vertical-cut proxy surfaces, the next step is to enable matrix-vector multiplication with the ``solution'' matrix $(\Iv-\Bv\Gv)^{-1}$, without actually computing and storing this large, dense matrix. The key, again, is to exploit the rank structure of the matrices involved. The hierarchical block matrix form of $\Iv - \Bv \Gv$ has the same HODLR structure as $\Gv$. After separating the solution matrix (TBD: OR ITS INVERSE?) into a product of the diagonal and off-diagonal parts of the matrix (for example, see Sec. 5.6 of \citeasnoun{martinsson2020fast}), the inverse of the product becomes the inverse of a block-diagonal matrix multiplied by the inverse of an identity-plus-low-rank matrix; the latter itself of identity-plus-low-rank form. Hence one can compute the inverses of the small blocks at the base of the hierarchy, then traverse the tree upward to build the matrices comprising the solution matrix, $(\Iv - \Bv\Gv)^{-1}$.

Since we have not formed the large, dense solution matrix itself, we then need an algorithm to compute the product of the solution matrix with a source (incident) field $\fv$. We can do this by reversing the order of operations in building the inverse matrices, and doing a downward traversal of the tree, from root to leaves, sequentially adding the products of the relevant matrices with $\fv$. Given the compression and proxy-surface techniques outlined above, instead of the usual $O(N^{3/2})$ complexity for general geometries, the total simulation complexity for metasurfaces is $O(N)$.

The collective procedure for a fast and accurate integral-equation-based metasurface simulation solver is given in \figref{fig2}. Starting from a square-grid discretization of a smoothed geometry (step 1), one adds quadrature-correction terms to the Green's-function matrix to enable higher-order accuracy (step 2). A binary tree data structure is then created for the hierchical compression and fast solution algorithms (step 3). As part of the compression, vertical-cut proxy surfaces are used, while random-matrix algorithms can be used for fast interpolative decomposition (ID) in steps 4 and 5. These matrices can then be used to represent the solution matrix, either accurately or with crude accuracy, where the latter is useful for preconditioning (steps 6,7). Finally, either the accurate inverse is multiplied by the excitation vector, or an iterative algorithm is used with the preconditioned system matrix (step 8). The orange boxes highlight the high-order quadrature ideas first proposed in \citeasnoun{duan2009high}, the green boxes highlight fast-direct solver ideas from \citeasnoun{gopal2022accelerated}, while the blue box highlights the metasurface-adapted proxy surface adaptations that we propose towards simulating, and designing, the largest possible metasurfaces. These steps are nearly independent of the dimensionality and polarization of the problem (2D TE, 2D TM, 3D), except that in the vector cases some reordering of the indices is helpful, and that the Green's function itself requires unique quadratures for each case. For the largest diameters, we use all of steps 1 through 8 for maximum computational efficiency, but one can use only a subset of the steps for smaller problems, to reduce implementation complexity.

\section{Computational examples}
\label{sec:Examples}
\begin{figure*}[htb]
  \includegraphics[width=\textwidth]{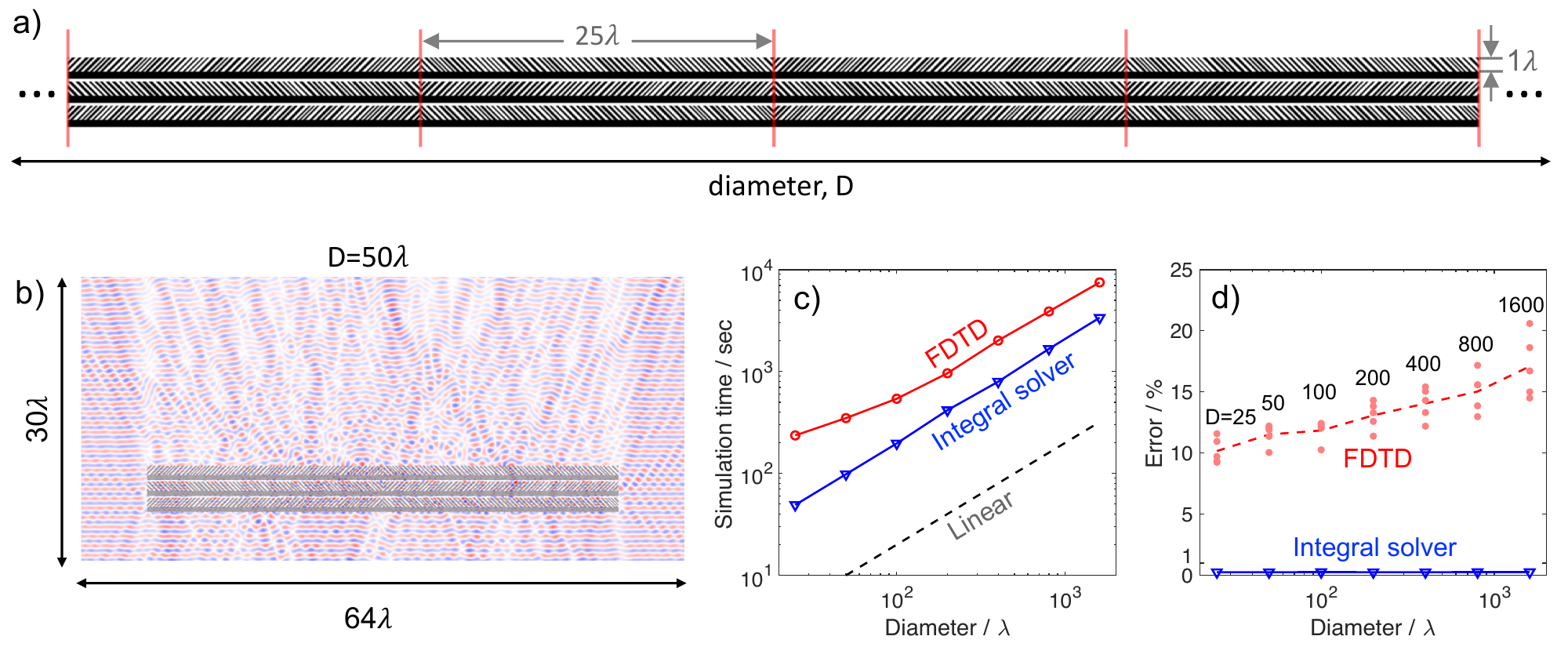}
  \caption{Comparison between the fast-direct integral-equation method and FDTD for nonlocal metasurfaces. (a) Geomety of the nonlocal metasurfaces: three layers, each generated from a non-symmetric center region comprising two blazed-grating regions (with randomly located slanted pillars) with opposite blazing directions, then adding the same the left/right-pattern outwards, consistently doubling the size. (b) Total electric field in our $50\lambda$-diameter metasurface. (c) Average simulation time over 5 metasurfaces at each diameter, with the simulation resolution fixed, which automatically leads to linear-in-time scaling for FDTD. The integral solver showing linear scaling is a demonstration of the predicted scaling from our compression and proxy-surface techniques. (d) Average simulation error for each technique. Whereas the FDTD solver average error increases from 10\% to 18\% as the diameter increases (with the largest single-simulation error exceeding 20\%), the integral solver maintains less than 1\% error.}
  \label{fig:fig3}
\end{figure*}

In this section, we demonstrate the capability of this fast-direct, linear-in-diameter algorithm to simulate large- and very-large-scale metasurfaces. We start with a prototypical problem of small to modest sizes, where we can compare the simulation times of the fast-direct integral solver with a finite-difference time-domain solver. The key result is that the FDTD simulation error increases steadily as the domain size increases, whereas the fast-direct solver error does not. Next, we turn our efforts to the \emph{design} of very-large-scale metasurfaces. First we design a metalens that is 20,000$\lambda$ in diameter, with a high efficiency even at a high numerical aperture. Finally, we show that this simulation approach works for any incident field, including in-plane waveguide-mode excitations, and design a large-area (1000$\lambda$-diameter) nonperiodic grating coupler, for large-area beam generation.

\subsection{Test case: fast-direct integral solver vs. FDTD}
As a first example, we compare the performance of the integral solver proposed above with \texttt{Meep}, a state-of-the-art FDTD solver~\cite{oskooi2010meep}. For benchmarking purposes, we consider 2D TE simulations (electric field out of plane) of the geometries in \figref{fig3}(a). These geometries constitute non-periodic, multilayered metasurfaces, generated by starting with a non-symmetric center region comprising two blazed-grating regions (with randomly situated slanted pillars) with opposite blazing directions, then mimicking the left/right-pattern outwards, consistently doubling the size. We use this structure to induce strong nonlocality (waves that propagate left-to-right in addition to upward/downward propagation). Metasurfaces in which the interaction is almost entirely local (predominantly up/down propagation) can already be simulated by techniques exploiting the locality~\cite{aieta2015multiwavelength,arbabi2015dielectric,khorasaninejad2016metalenses,wang2018broadband,pestourie2018inverse,lin2019overlapping}, but such metasurfaces have limitations in performance if they cannot take advantage of nonlocality~\cite{shastri2023nonlocal,chen2021dielectric}. A primary goal in the metasurface community is in taking steps to nonlocal~\cite{overvig2022diffractive,goh2022nonlocal,overvig2021thermal,kwon2020dual,kwon2018nonlocal} and multilayer~\cite{arbabi2017planar,mansouree2020multifunctional,arbabi2018mems,kwon2020single,arbabi2016miniature,zhou2018multilayer,christiansen2020fullwave} metasurfaces, which \figref{fig3}(a) effects. For generic wavelength $\lambda$, we consider 7 multilayer designs: the first has diameter $50\lambda$, the second $100\lambda$, and so forth (doubling each time), up to $25\times 2^6 \lambda = 1600\lambda$ diameter. The grating thickness is taken to be $1\lambda$, the substrate $0.45\lambda$, and the air gaps (mimicking spacer layers~\cite{zhou2018multilayer,arbabi2018mems,kwon2020single,christiansen2020fullwave}) $0.2\lambda$. The scatterer material is taken to have refractive index of 2, typical of transparent materials at visible frequencies (e.g. SiN~\cite{beliaev2022optical}). For each diameter we create 5 sample metasurfaces to simulate (with different random pillar positions), to smooth the error rates. The resolution of the FDTD solver is taken to be 80 points per free-space wavelength (chosen to ensure no greater than 10\% error at the smallest diameter), while the fast-direct integral solver uses 40 points per wavelength. To compute error rates, we find the ``ground truth'' solution at each diameter by using the fast-direct solver at high enough resolution to ensure converge to $\lesssim 0.1\%$ accuracy. We also verify for the smallest diameter structure that the FDTD solutions indeed converge (albeit slowly) to the ground-truth solution.

\Figref{fig3}(b-d) depict data from the fast-direct integral-equation solver and the FDTD solver for various scatterer diameters. \figref{fig3}(b) shows the total electric field (from our fast-direct solver) in the smallest-diameter case considered, with $50\lambda$ diameter. The image shows the complex interference patterns common to nonlocal metasurfaces. \figref{fig3}(c) shows the simulation times for both solvers as a function of the diameter of the scatterer. The number of grid points per wavelength is kept fixed for each simulation, and the FDTD termination criterion measures the field energy in the scattering body. The absolute value of the simulation time is not crucial: the fast-direct solver is about twice as fast, but these numbers can be scaled based on CPU/GPU implementations, number of parallel cores, etc. (The integral-equation solver is run on a high-memory, 8-node, 3.3 GHz compute node, while the FDTD solver is run on a 24-cpu, 2.9 GHz compute node.) The total simulation time for a given application will also highly depend on whether a multi-frequency simulation is required (where FDTD can be very fast), multi-frequency design is required (where FDTD is less fast due to slow adjoint computations~\cite{hammond2022high}), multi-angle simulations or designs (where the integral-equation solver can be very fast), multi-iteration designs (where the integral-equation solver has additional speedups), and so forth. In the quest for pushing simulations to the largest size scales, the key question is the amount of time required for accurate simulation, as a function of scatterer size.

As seen in \figref{fig3}(c), both solvers have simulation times that scale linearly with the size of the solver. This is almost a trivial fact for the FDTD solver, since we keep the resolution fixed, and the number of time steps hardly changes. For the fast-direct integral-equation solver, however, the linear scaling is a significant achievement. This shows that our implementation of dense-matrix compression as well as hierarchical simulation algorithms have indeed correctly led to linear-in-time, integral-equation-based simulation. 

The payoff for the linear-in-time integral-equation solver is shown in \figref{fig3}(d). The FDTD solver error increases as the size of the simulation region increases, as expected for a finite-difference discretization of the poorly conditioned differential form of Maxwell's equations. One can see that the possibly tolerable 10\% errors at the smaller diameters are approaching 20\% at the $1600\lambda$-diameter structure. At the extraordinarily large sizes that we can consider in the next section, the errors are too large to be useful, and increasing the resolution to the point of bringing the errors back to say $5\%$ would incur enormous increases in simulation time. By contrast, the matrices created by discretizing the integral equations are well-conditioned, implying the possibility for high accuracy at very large sizes, which is born out in the integral-solver data in \figref{fig3}(d). The fast-direct solver can maintain accuracies better than 1\%, which is important not only for simulation purposes, but additionally for iterative design algorithms, where errors in consecutive geometries can hamper the convergence speed of an optimization algorithm itself.

\subsection{Fullwave inverse design of a centimeter-scale metalens}
\begin{figure*}[htb]
  \includegraphics[width=\textwidth]{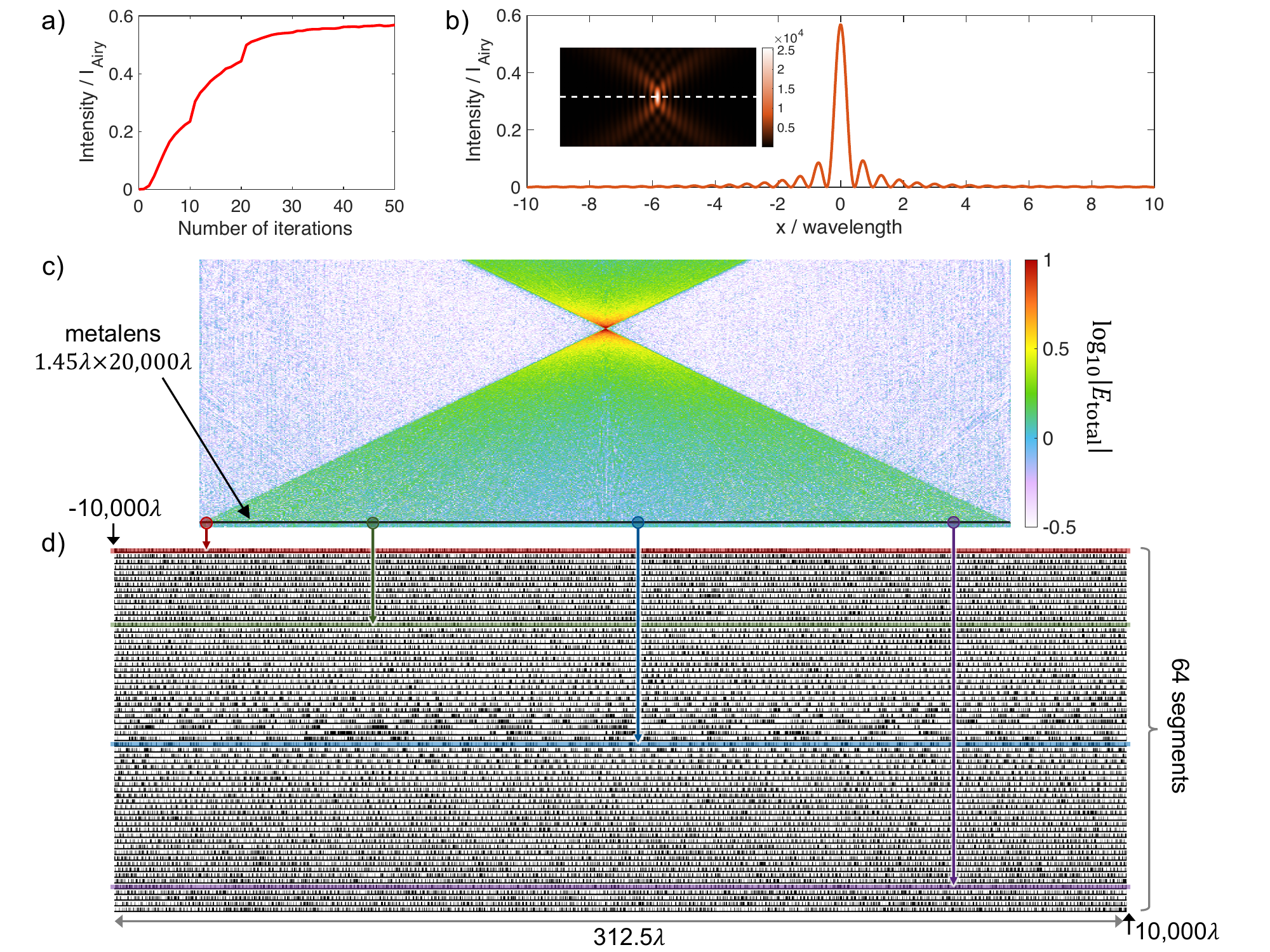}
  \caption{Optimized 2D metalens with a 20,000$\lambda$ diameter. (a) Evolution of the figure of merit over 50 optimization iterations, requiring less than three days to complete on a 16-CPU cluster. The final Strehl ratio reaches 56\%. (b) Electric field intensity distribution on the focal plane. Inset: field intensity map near the focal plane (white dashed line), in a region $10\lambda\times20\lambda$. (c) Electric field from the metalens through the focal plane. The absolute field value is plotted in log scale, with the colormap floor restricted to $10^{-0.5}$ to de-emphasize the fields within the scatterer itself. The location of the $1.45\lambda\times20,000\lambda$ metalens is represented by the black line. (d) Detailed geometric structure of the metalens. Due to the large aspect ratio, the metalens is divided into 64 vertically stacked segments, each of which is $312.5\lambda$ wide. Shaded portions in (c) are mapped to their corresponding rows in (d) to illustrate the segmentation.}
  \label{fig:fig4}
\end{figure*}

Next we use the fast-direct integral-equation solver to design metasurfaces at extreme size scales. We consider a cylindrical metalens with a refractive index of 2 (as for SiN in the visible~\cite{beliaev2022optical}), under TE-polarized illumination (scalar electric field), with a diameter that is 20,000 wavelengths in size. This size corresponds to a \SI{1}{cm} diameter for \SI{500}{nm} wavelength. To the best of our knowledge, this is significantly larger than the largest fullwave metalens designs to date~\cite{christiansen2020fullwave}, which had diameters of 129$\lambda$ (for single-wavelength thickness) or 100$\lambda$ (for 10$\lambda$ thickness). We design for a numerical aperture of 0.9, a high NA that requires strong light focusing, and where unit-cell and related approximations break down. We consider a metasurface of ``pillars'' $1\lambda$ in height, with arbitrary widths and separations, which together comprise about 200,000 design variables (the total number is not fixed as pillars can be created or destroyed during the optimization process). We use adjoint-based inverse design~\cite{Jensen2011,Miller2012,piggott2015inverse} to compute the gradients with respect to all degrees of freedom at every iteration, and we use gradient ascent for the optimization algorithm. We also impose a minimum-feature-size constraint by forbidding widths smaller than $\lambda/50$ (\SI{10}{nm} for \SI{500}{nm} wavelength). To balance the tradeoff between speed and accuracy, we use a simulation grid spacing of $\lambda/36$, which corresponds to field errors at a reasonable 2--3\%. The initial compression of $\Gv$ requires 230 seconds, but can be reused for every iteration. Within every iteration, building $(\Iv-\Bv\Gv)^{-1}$ takes 2330 seconds, and the solution via BICGSTAB takes 1150 sec for both direct and adjoint simulations, for a total time per iteration of 4630 seconds.

\Figref{fig4}(a) shows the evolution of the focal-point intensity of the metalens over the course of 50 optimization steps. By the end of the optimization, the Strehl ratio (defined as the focal-point intensity as a fraction of an ideal Airy-beam focusing intensity) reaches 56\%, quite a high efficiency consider the numerical aperture of 0.9. This Strehl ratio is defined relative to \emph{all power} incident upon the metasurface, not only the power transmitted through the surface. \Figref{fig4}(b,c) show the fields of the optimal design: one can clearly see nearly ideal performance in the focal plane (\figref{fig4}(b)), and the excellent focusing performance (\figref{fig4}(c)). \Figref{fig4}(d) shows the full metalens design. Given the difficulty of visualizing a structure with an approximately 20,000:1.5 aspect ratio, we partition the metasurface into 64 segments, each $312.5\lambda$ wide, and vertically stack the neighboring segments. The shaded circles in \figref{fig4}(c) map different segments within the metasurface to the corresponding design for that segment in \figref{fig4}(d). The entire fullwave optimization process for the 20,000$\lambda$ diameter metasurface requires only 64 hours to complete on a single 16-CPU compute node cluster. Further speedups should be possible with a more extensive parallelization effort.

\begin{figure*}[htb]
  \includegraphics[width=\textwidth]{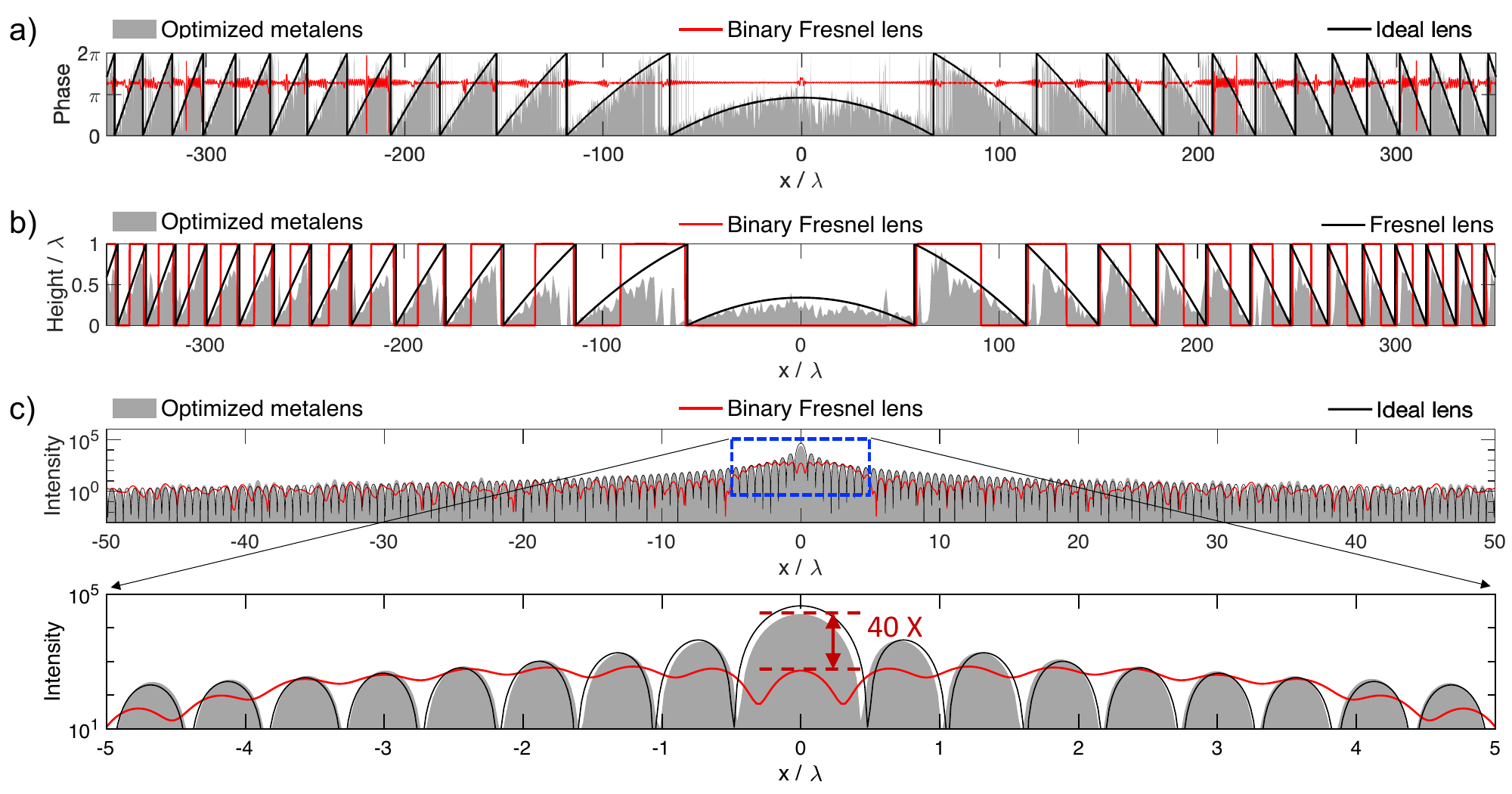}
  \caption{(a) Phase performance of the ideal lens (black), optimized metalens (gray), and a binary Fresnel lens (red). (b) The optimized metalens geometry shows an average-height profile similar to a Fresnel lens (averaged over one-micron regions), albeit in a lithographic form factor and with significant corrections. (c) Intensity enhancements of the binary Fresnel lens versus the optimized metalens.}
  \label{fig:fig6}
\end{figure*}

Inverse design offers little guidance as to why an optimized structure performs well. In \figref{fig6} we present a partial mechanism for understanding the optimized design. First, \figref{fig6}(a) shows that the optimized lens effectivelly creates the correct outgoing-field phase profile across the surface of the metalens, even at the extreme edges where the phase varies more rapidly. Then, in \figref{fig6}(b) we see that a locally averaged (over every one micron) height profile of the optimized metalens resembles that of a Fresnel lens, albeit with some significant modifications near the height discontinuities, presumably to smooth and enhance the performance. A binary Fresnel lens, depicted in (red) in \figref{fig6}(b), does not create an effective phase profile in \figref{fig6}(a), which leads to subpar intensity focusing as depicted in \figref{fig6}(c). The optimized metalens offers a 40X relative intensity enhancement. Hence, we can interpret the complex structural patterns of the optimized metalens as a device that creates the ideal outgoing-field phase profile, while accounting for all multiple-scattering effects present in such strong-field-bending scenarios.

\subsection{Fullwave inverse design of a millimeter-scale waveguide coupler}
\begin{figure*}[htb]
  \includegraphics[width=\textwidth]{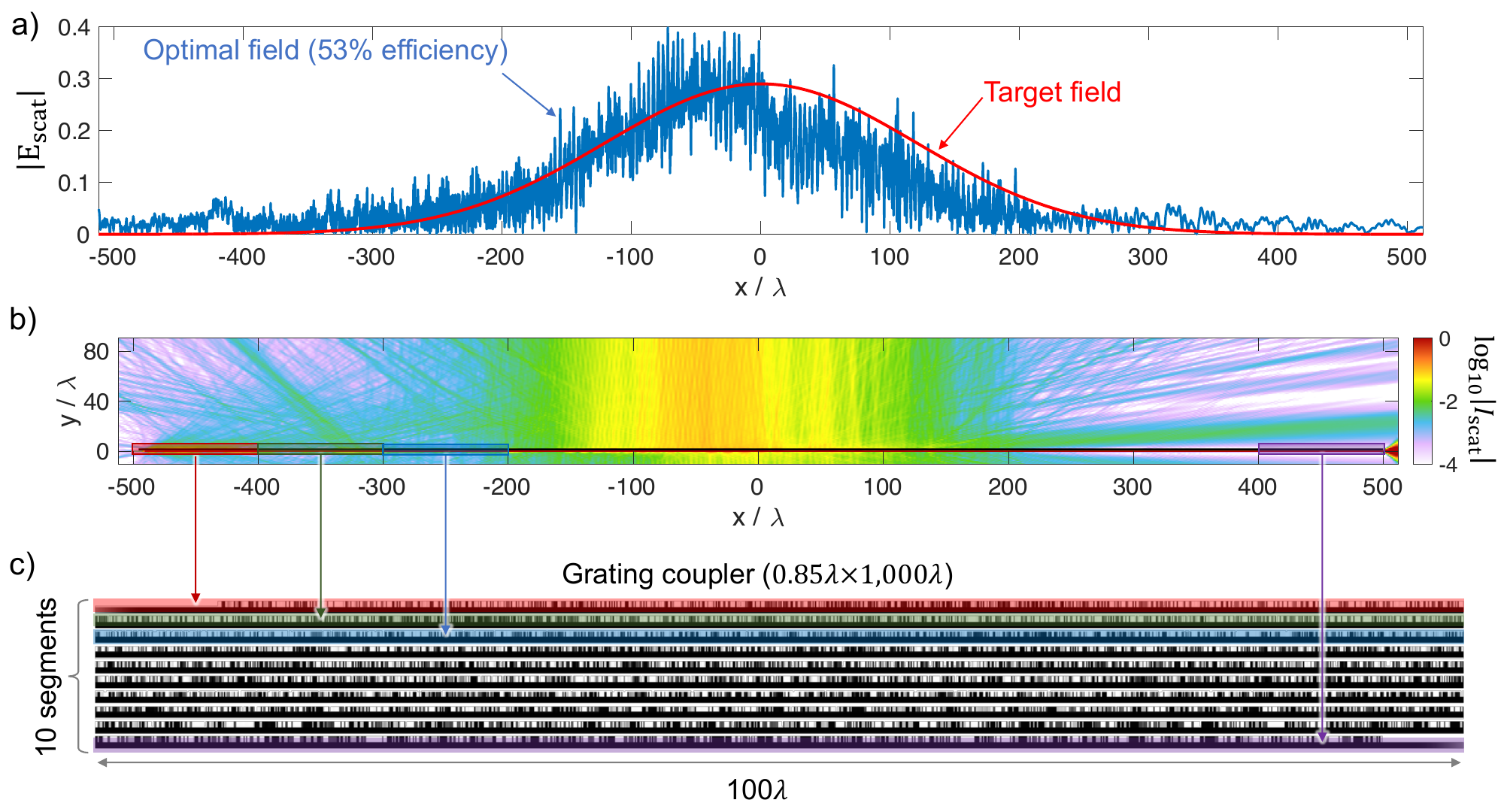}
  \caption{$1,000\lambda$-long waveguide coupler inverse designed via 750 iterations, each with two full simulations by our fast-direct solver. (a) Target (red) and optimal (blue) scattered field on a horizontal plane $90\lambda$ above the coupler, excited from the left by the primary TE mode of the waveguide. The coupling efficiency is 53\%, defined as the power flow into the target mode. (b) Field intensity, showing strong coupling to a vertically propagating Gaussian mode. The primary loss mechanism is downward scattering, which is likely unavoidable given the refractive indices and design constraints (height, lithography, etc.). (c) Depiction of the optimal metalens, partitioned into 10 segments of 100$\lambda$ widths.}
  \label{fig:fig_waveguide}
\end{figure*}
Large-area, nonperiodic metasurfaces can also be excited from the side, scattering waves into free space, thereby acting as grating couplers. Such side-coupled excitations experience very long propagation lengths, rendering typical unit-cell approximations~\cite{mansouree2020multifunctional} ineffective or of modest efficiency~\cite{khorasaninejad2015achromatic,wang2018broadband,chen2018broadband,chung2020high}. The fast-direct solver approach is effective for any excitation. Hence the same algorithms  and solver described above for conventional metasurface applications can seamlessly be applied to grating couplers and related beam-coupling devices. In this section, we design a 1000-wavelength-long grating coupler, well beyond the limits of conventional finite-difference and finite-element solvers, for coupling a waveguide mode to a large-area Gaussian beam. 

For a telecommunications wavelength of \SI{1550}{nm}, the coupler length of 1000 wavelengths corresponds to a physical length of \SI{1.55}{mm}. We consider a material with permittivity 2.2, close to that of ${\rm SiO_2}$~\cite{arosa2020refractive}, and TE-polarized propagation. The grating is taken to comprise pillars of $0.4\lambda$ height on a $0.45\lambda$-thick substrate. We use the same shape-optimization algorithm described above, utilizing adjoint gradients at each iteration of the optimization, to maximize the out-coupling efficiency to a target Gaussian beam with field pattern $E = E_0 \exp{\left(-x^2/2w^2\right)}$, with parameter $w$ set to 341$\lambda$, creating a full-width at half-maximum of about $800\lambda$, corresponding to a \SI{1.24}{mm} beam size for $\lambda = \SI{1550}{nm}$. The amplitude $E_0$ is chosen to correspond to all incoming waveguide power being scattered into the Gaussian mode. The waveguide mode is excited by an electric dipole source located multiple wavelengths from the start of the design region, to prevent back-reflection of the source near field. 

To balance the tradeoff between simulation speed and accuracy, we use resolution $\lambda/40$, for which we find errors of apprximately 5\%. The compression stage takes 30 seconds, which can be reused for all simulations during the optimization. Then simulating the forward and adjoint excitations takes a little more than two minutes per iteration, so that a 750-iteration optimization on an 8-core, 2.9 GHz compute node can be completed in about 33 hours. Starting from a random initial pattern, the design process iterates to a grating coupler with 53\% coupling efficiency to a large-area Gaussian beam.

The final design and its corresponding field patterns are shown in \figref{fig_waveguide}. \Figref{fig_waveguide}(a) shows the target field (red) alongside the simulated field of the optimal design (blue), along a plane 90$\lambda$ above the grating-coupler surface. The full spatial distribution is shown in \figref{fig_waveguide}(b), while the optimal design is shown in \figref{fig_waveguide}(c). (For visualization purposes, the intensity is shown in log scale, with small-scale oscillations filtered out by Gaussian convolution. \Figref{fig_waveguide}(a) clearly depicts the oscillations. The colormap ceiling is saturated to de-emphasize the fields within the scatterer.) To accommodate the high aspect ratio of the device we split it into ten segments and arrange them vertically. From \figref{fig_waveguide}(a,b), one can see that the optimal field closely approaches the target. The key factor inhibiting even higher efficiency is power that is scattered downwards, due to the symmetric air regions above and below the device (pattern and thin substrate). For example, in the unpatterned structure, a dipole source placed directly above the waveguide radiates nearly symmetrically in the up and down directions: approximately 50--55\% upwards, and 45--50\% downwards. An examination of the polarization fields induced in the optimal design shows little vertical phase variation within the grating-coupler pillars, suggesting that the pillars cannot effectively cancel the downwards radiation. Hence, the optimal design is closely approaching the upper limit of what is possible for a grating coupler in this architecture.

\section{Conclusion and outlook}
\label{sec:Conclusion}
In this work, we propose a fast-direct integral-equation solver for large-scale two-dimensional metasurface simulations. The solver takes advantage of high-order quadrature schemes to attain high simulation accuracy for modest resolution, as well as tailored algorithmic improvements within the emerging fast-direct solver framework~\cite{martinsson2007fast,duan2009high,martinsson2020fast,gopal2022accelerated}, to ultimately achieve linear-in-time scaling of the computations as a function of metasurface diameter. Our simulations show significant scalability advantages relative to finite-difference simulations, offering the possibility to reach extreme sizes and scales without sacrificing any aspects of wave scattering. We realize these possibilities with two optimal designs; first, for a high-NA metalens with $20,000\lambda$ diameter (corresponding to a cm-scale design for \SI{500}{nm} wavelength), then, for a high-efficiency grating coupler with a 1000$\lambda$ design-region length, targeting large-area beam generation applications.

The key next steps are the translation of these design principles to three dimensions. Two-dimensional simulations with rotational symmetries would already be of interest for many three-dimensional applications with cylindrical symmetry (e.g. \citeasnoun{christiansen2020fullwave}), after which the three-dimensional case follows. Each of these cases requires significant numerics and code development, as the Green's functions are different in each scenario. One interesting potential tradeoff in three dimensions would be to consider a surface-integral-equation approach, which has more complex quadrature implementation requirements, but which may yield computational speed advantages. As these simulations tools are developed, one can expect that fast-direct integral-equation solvers will unlock the possibility to do nanophotonic design at the scale of conventional optics.

\section{Acknowledgements}
W. Xue, H. Zhang and O.D. Miller were partially supported by the Air Force Office of Scientific Research Grant No. FA9550-22-1-0393. H. Zhang, A. Gopal, and V. Rokhlin were supported by ONR grant N00014-18-1-2353 and by NSF grant DMS-195275. 

\bibliography{bib}
\end{document}